\documentclass[conference]{IEEEtran}
\IEEEoverridecommandlockouts

\usepackage{cite}
\usepackage[hidelinks]{hyperref}
\usepackage{amsmath,amssymb,amsfonts}
\usepackage{algorithmic}
\usepackage{graphicx}
\usepackage{textcomp}
\usepackage{booktabs}
\usepackage{xcolor}
\def\BibTeX{{\rm B\kern-.05em{\sc i\kern-.025em b}\kern-.08em
    T\kern-.1667em\lower.7ex\hbox{E}\kern-.125emX}}
\begin{document}

\title{Context-Aware Markov VAE for CSI Compression in Wireless Systems\\
\thanks{This work was supported by Ericsson Research and the Wallenberg AI, Autonomous Systems, and Software Program (WASP) funded by the Knut and Alice Wallenberg Foundation. The work of N. Pappas has been supported in part by ELLIIT and the European Union (6G-LEADER, 101192080).}
}

\author{\IEEEauthorblockN{Efstathios Chatziloizos\IEEEauthorrefmark{1}\IEEEauthorrefmark{2}, Konstantinos Vandikas\IEEEauthorrefmark{1}, Aneta Vulgarakis Feljan\IEEEauthorrefmark{1}, Zheng Chen\IEEEauthorrefmark{3}, and Nikolaos Pappas\IEEEauthorrefmark{2}}
\IEEEauthorblockA{\IEEEauthorrefmark{1} Ericsson Research, Stockholm, Sweden}
\IEEEauthorblockA{\IEEEauthorrefmark{2} 
Department of Computer and Information Science, Linköping University, Linköping, Sweden}
\IEEEauthorblockA{\IEEEauthorrefmark{3} Department of Electrical Engineering, Linköping University, Linköping, Sweden}
Email: {\{efstathios.chatziloizos, konstantinos.vandikas, aneta.vulgarakis\}}@ericsson.com, \\ {\{efstathios.chatziloizos, zheng.chen, nikolaos.pappas}\}@liu.se}

\maketitle

\begin{abstract}
This paper considers neural channel state information (CSI) compression for time-varying massive multiple-input multiple-output (MIMO) channels in frequency division duplex (FDD) systems with limited feedback resources. The main challenge lies in obtaining a compact and efficient representation of the CSI 
given that it exhibits strong temporal correlation across successive snapshots. 
Existing memoryless compression models do not exploit this property, while simple temporal extensions often incorporate multiple observations without explicitly modeling the latent dynamics. We propose a context-aware compression framework based on a $k$-memory Markov variational autoencoder ($k$-MMVAE), which uses a finite temporal window to capture the evolution of CSI in the latent space. The model introduces Markov-structured latent dynamics with finite memory, enabling efficient use of temporal dependencies for compression. Simulation results show that the proposed approach improves target CSI reconstruction performance compared to memoryless and weakly sequential baselines, particularly at low and moderate compression rates. These results suggest that explicit latent temporal modeling can provide an effective mechanism for CSI compression under limited feedback constraints.
\end{abstract}


\section{Introduction}

Massive multiple-input multiple-output (MIMO) is a key technology for achieving high spectral efficiency and link reliability in modern wireless systems \cite{larrson2014, fundamentals}. 
Channel state information (CSI) is essential for coherent operation in massive MIMO, as accurate CSI enables phase-aligned beamforming across many antennas.
In frequency division duplex (FDD) operation, the base station (BS) transmits downlink pilots to enable channel estimation at the user equipment (UE), after which the estimated CSI is fed back to the BS before downlink data transmission. This incurs a feedback overhead that scales with the number of BS antennas \cite{love2008}, motivating the need for efficient CSI compression \cite{kuo2012cs}.

A broad range of CSI feedback compression/reduction methods has been studied. Early approaches exploited sparsity via codebook design or compressive sensing \cite{kuo2012cs,rao2014distributed}. Later on, deep learning methods have been considered to learn compact latent representations directly from data. In particular, CsiNet established the autoencoder paradigm for CSI feedback \cite{wen2018csinet}. Subsequent works extended this approach by incorporating temporal correlation, multi-resolution architectures, and rate--distortion-aware compression \cite{wang2019csinetlstm,lu2020crnet,mashhadi2021deepcmc}. Recent work has further explored attention-based and transformer-based architectures, as well as adaptive and data-efficient feedback schemes, to exploit temporal structure better and improve generalization across environments \cite{cui2022transnet,bi2022transformer,liu2024efficient,shen2024bilstm}.

Despite the progress in this research direction, limitations remain. Compressive sensing methods typically rely on iterative reconstruction and can be computationally demanding at high compression ratios \cite{kuo2012cs,rao2014distributed}. Deep learning approaches are often memoryless or implicitly incorporate temporal context, without explicitly modeling latent dynamics \cite{wen2018csinet,wang2019csinetlstm}. Consequently, they may not fully exploit the structure of time-varying CSI, particularly when the objective is to accurately reconstruct the latest snapshot with limited feedback and reduced latent dimensionality. This limitation becomes more critical when the temporal context is partially unavailable due to missing or delayed feedback, requiring additional designs that are robust to incomplete observations.

In this paper, we propose a context-aware CSI compression design based on a $k$-memory Markov variational autoencoder ($k$-MMVAE), where $k$ denotes the number of past time steps retained in the temporal context window. The model introduces Markov-structured latent evolution with finite memory to capture temporal dependencies across CSI snapshots. We investigate whether explicit latent temporal modeling provides an effective compression mechanism for target CSI reconstruction under limited feedback, supported by a systematic comparison with memoryless and weakly sequential baselines.

\section{System Model}

We consider an FDD massive MIMO system where downlink CSI is estimated at the UE and fed back to the BS over a limited uplink channel. Due to the high dimensionality of CSI in massive MIMO systems, direct feedback of the full CSI vector is impractical, motivating the need for efficient compression schemes.

Let $\mathbf{x}_t \in \mathbb{R}^{N}$ denote the CSI vector at time step $t$, where $N = 2R$ corresponds to the real and imaginary components of the channel across $R$ antennas. CSI evolves smoothly over time due to user mobility and environmental dynamics. We therefore consider CSI trajectories as sequences of vectors
\begin{equation}
\mathbf{x}_{1:T} = (\mathbf{x}_1, \mathbf{x}_2, \dots, \mathbf{x}_T), \quad \mathbf{x}_t \in \mathbb{R}^{N}, \; t = 1,\dots,T.
\end{equation}

\begin{figure}[t]
    \centering
    \includegraphics[width=0.99\columnwidth]{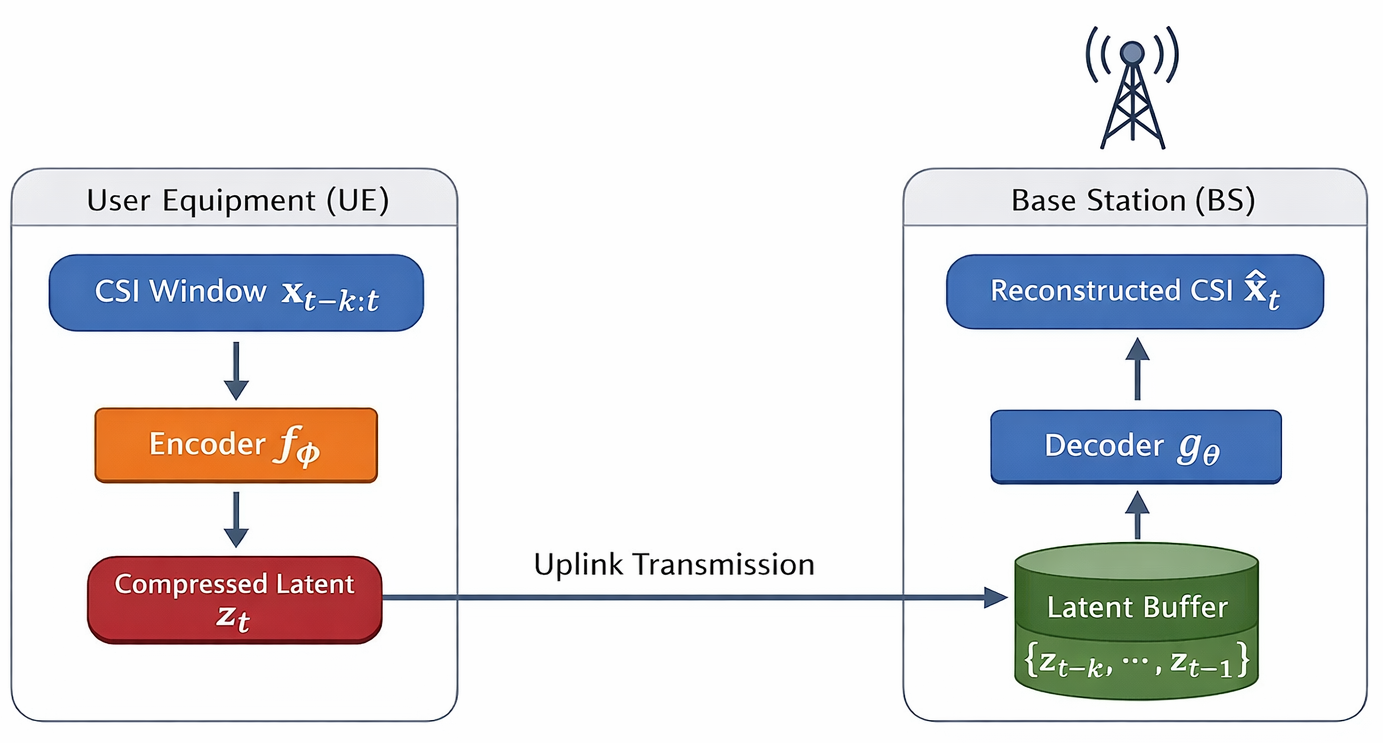}
    \caption{End-to-end CSI feedback pipeline. At each time step, the UE encodes the available CSI context into a latent representation, transmits the latent over the uplink, and the BS reconstructs the CSI using the received latent variables and any stored latent context.}
    \label{fig:pipeline}
\end{figure}

We consider a sequential communication pipeline illustrated in Fig.~\ref{fig:pipeline}. At each time step $t$, the UE observes the current CSI, along with any available past observations, and compresses it into a low-dimensional latent representation. This latent representation is transmitted to the BS over the uplink channel under communication constraints.

At the BS, the received latent variables are used to reconstruct the CSI. Importantly, the decoder has access only to the received latent variables and any previously stored latents, but not to the original CSI observations. For models with temporal structure, previously received latent variables are stored and reused during reconstruction.

The compression process is modeled by an encoder $\mathbf{z}_t = f_\phi(\mathcal{C}_t)$, where $\mathbf{z}_t \in \mathbb{R}^d$ with $d \ll N$, and $\mathcal{C}_t$ denotes the model-dependent context. The compression ratio is defined by the relative dimensionality reduction $d/N$, where smaller values correspond to stronger compression and typically result in higher reconstruction distortion due to the limited representational capacity of the latent space. This context may consist of the current snapshot $\mathbf{x}_t$, a temporal window $\mathbf{x}_{t-k:t}$ of length $k+1$, previous latent variables such as $\mathbf{z}_{t-1}$, or any combination of these.

At the receiver, the CSI is reconstructed using a decoder
$\hat{\mathbf{x}}_t = g_\theta(\mathcal{Z}_t)$, where $\mathcal{Z}_t$ denotes the set of latent variables available at the BS, which may consist of $\mathbf{z}_t$, a window of latent vectors $\mathbf{z}_{t-k:t}$ of length $k+1$, or any subset of previously transmitted latents depending on the model.

In practice, the encoder may not always have access to a complete history of past CSI observations. This can occur, for example, due to missing or delayed pilot measurements, interrupted feedback updates, or outdated CSI acquisition in dynamic wireless environments. We therefore also consider settings in which part of the past CSI context is unavailable at the encoder input. Studying this regime is important because temporal compression methods rely on past observations, and their practical usefulness depends not only on their peak reconstruction performance but also on their robustness to incomplete temporal context, as evaluated in Section~\ref{sec:sim_results}.

The compression performance is characterized using normalized mean squared error (NMSE), measured in dB
\begin{equation}
\text{NMSE} = \frac{\|\mathbf{x}_t - \hat{\mathbf{x}}_t\|^2}{\|\mathbf{x}_t\|^2}, \text{ NMSE}_{\text{dB}} = 10 \log_{10}(\text{NMSE}).
\end{equation}

\section{Compression Models}

\subsection{Baseline Models}

We consider several baseline architectures with increasing use of temporal context.

\textbf{AE/VAE:}
Memoryless models that compress a single CSI snapshot. The AE uses a deterministic latent representation~\cite{hinton2006ae}, while the VAE introduces a Gaussian latent variable~\cite{kingma2013vae}. The encoding and decoding processes can be abstractly written as
\begin{equation}
\mathbf{z} = f_\phi(\mathbf{x}_t), \qquad \hat{\mathbf{x}}_t = g_\theta(\mathbf{z}).
\end{equation}

\textbf{CVAE:}
Encodes the entire trajectory into a single latent variable
\begin{equation}
q_\phi(\mathbf{z} \mid \mathbf{x}_{1:T}),
\end{equation}
from which either the full trajectory or only the final snapshot is reconstructed.

\textbf{TSVAE:}
A sequential baseline where each latent depends on a local observation window
\begin{equation}
q_\phi(\mathbf{z}_t \mid \mathbf{x}_{t-k:t}), \qquad \hat{\mathbf{x}}_t = g_\theta(\mathbf{z}_t).
\end{equation}
Temporal context is used at the encoder, but latent variables remain independent under the prior.

Overall, these baselines range from memoryless compression (AE/VAE), to global temporal compression (CVAE), to locally conditioned sequential models without latent dynamics (TSVAE).

\subsection{$k$-Memory Markov VAE ($k$-MMVAE)}

\begin{figure*}[t]
    \centering
    \includegraphics[width=0.8\textwidth]{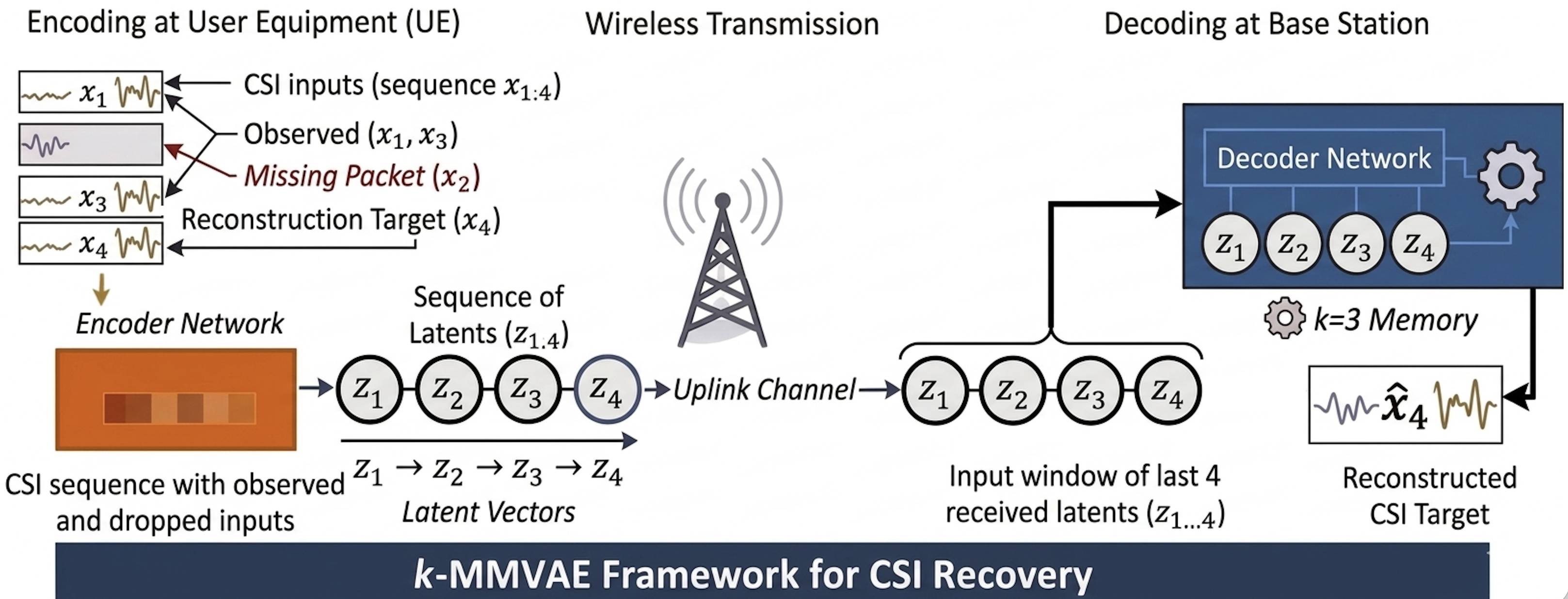}
    \caption{$k$-MMVAE-based CSI feedback framework. The UE encodes CSI observations into latent variables, which are transmitted over the uplink. The BS reconstructs the CSI using a finite memory of past latents. Missing packets correspond to unavailable past CSI observations at the encoder input.}
    \label{fig:kmmvae}
\end{figure*}

The $k$-MMVAE constitutes the proposed approach of this work, and its operation within the CSI feedback pipeline is illustrated in Fig.~\ref{fig:kmmvae}. The model is adapted from~\cite{bock2022tvvae}, where a similar variational framework is used to estimate time-varying channels and is applied here for CSI feedback compression. Here, $k$ denotes the memory length, i.e., the number of past time steps retained in the encoder and decoder context window.

The latent variables follow a first-order Markov prior
\begin{equation}
p(\mathbf{z}_{1:T}) = \prod_{t=1}^{T} p(\mathbf{z}_t \mid \mathbf{z}_{t-1}),
\end{equation}
while the approximate posterior incorporates both the previous latent and a finite observation window
\begin{equation}
q_\phi(\mathbf{z}_{1:T} \mid \mathbf{x}_{1:T})
=
\prod_{t=1}^{T}
q_\phi(\mathbf{z}_t \mid \mathbf{z}_{t-1}, \mathbf{x}_{t-k:t}).
\end{equation}
The decoder reconstructs the CSI from a latent window
\begin{equation}
\hat{\mathbf{x}}_t = g_\theta(\mathbf{z}_{t-k:t}),
\end{equation}
allowing temporal dependencies to be represented explicitly in the latent space.

The training objective is
\begin{equation}
\begin{aligned}
\mathcal{L}_{k\text{-}MMVAE}
&=
\frac{1}{T}\sum_{t=1}^{T}
\Big[
\|\mathbf{x}_t-\hat{\mathbf{x}}_t\|^2 \\
&+
\beta \, D_{KL}\!\left(
q_\phi(\mathbf{z}_t \mid \mathbf{z}_{t-1}, \mathbf{x}_{t-k:t})
\,\|\, 
p(\mathbf{z}_t \mid \mathbf{z}_{t-1})
\right)
\Big].
\end{aligned}
\end{equation}

In the target-only setting, only the final snapshot is reconstructed, while KL regularization is applied across all time steps. Compared with~\cite{bock2022tvvae}, we use a deterministic MSE reconstruction loss.

Unlike other baselines, $k$-MMVAE explicitly models temporal evolution in the latent space via a Markov prior while leveraging finite memory at both the encoder and the decoder.

\section{Dataset and Spatial Correlation Analysis}

The dataset is generated using the QuaDRiGa channel simulator \cite{quadriga}, following a geometry-based stochastic channel modeling approach. We consider an urban macrocell (UMa) scenario based on the 3GPP 38.901 model, closely following the data generation procedure in~\cite{bock2022tvvae}.
The carrier frequency is $f_c = 2.1$ GHz, with a snapshot interval of $T_s = 0.5$ ms. The base station is equipped with a uniform linear array (ULA) of $R = 32$ antennas with inter-element spacing $\lambda/2$, where $\lambda$ denotes the carrier wavelength.

Each CSI trajectory consists of $T = 8$ snapshots (subsampled from longer trajectories when needed). User mobility is modeled via a Rayleigh-distributed speed $v \sim \text{Rayleigh}(\sigma = 2)$ (corresponding to $\sigma^2 = 4$). Propagation conditions include a mixture of line-of-sight (LOS) and non-line-of-sight (NLOS) scenarios, with $80\%$ of users randomly assigned to indoor environments.
These settings ensure realistic spatial and temporal correlation properties of the generated CSI trajectories.


For each trajectory, multipath channel coefficients are generated using QuaDRiGa. The narrowband channel vector at time $t$ is obtained as
\begin{equation}
\mathbf{h}_t = \sum_{l=1}^{L} a_l(t) e^{-j 2 \pi f_c \tau_l(t)},
\end{equation}
where $a_l(t)$ and $\tau_l(t)$ denote the gain and delay of path $l$.
Each trajectory is normalized to remove large-scale fading variations, i.e.,
$\mathbf{h}_t \leftarrow \frac{\mathbf{h}_t}{\sqrt{\mathbb{E}[\|\mathbf{h}_t\|^2]}}$.
The final dataset consists of $100,000$ training trajectories, $10,000$ validation trajectories, and $10,000$ test trajectories.


Before evaluating temporal compression methods, we examine the spatial structure of the dataset to understand the inherent redundancy. More explicitly, we consider the spatial covariance matrix
$\mathbf{C} = \mathbb{E}[\mathbf{h}\mathbf{h}^H]$,
estimated empirically over the dataset.
To remove scaling effects, we compute the correlation matrix
\begin{equation}
\rho_{ij} = \frac{C_{ij}}{\sqrt{C_{ii}C_{jj}}}.
\end{equation}
For the uniform linear array (ULA), correlation primarily depends on antenna separation. Let us define
\begin{equation}
\rho(d) = \frac{1}{R-d} \sum_{i=1}^{R-d} |\rho_{i,i+d}|.
\end{equation}
We observe strong short-range correlation with $\rho(1) \approx 0.69$ and $\rho(2) \approx 0.52$, followed by gradual decay.


After computing the sample covariance matrix, we can analyze the eigenvalue spectrum of $\mathbf{C}$ and compute the effective rank $
r_{\text{eff}} = \exp\left(-\sum_i p_i \log p_i \right)$,
where $p_i$ are normalized eigenvalues. We obtain $r_{\text{eff}} \approx 14.75$, significantly lower than $R=32$, indicating strong spatial redundancy.


These results indicate that \textit{CSI lies in a low-dimensional spatial subspace, such that memoryless compression already exploits a large portion of the redundancy}. Consequently, the gains from temporal modeling should be understood as an additional source of improvement in terms of communication reduction. 

\section{Simulation Results}
All models share the same neural backbone with comparable capacity to ensure a fair comparison. They are trained using the Adam optimizer with a learning rate of $5\times 10^{-4}$ and a batch size of 256 for up to 80 epochs, with early stopping (patience 15) based on validation loss. 

All models are trained using a $\beta$-weighted variational objective that balances reconstruction accuracy and latent regularization via the KL divergence. For sequential models, the loss is averaged over time steps. We set $\beta = 3 \times 10^{-6}$ in all experiments.

\label{sec:sim_results}

We evaluate models under two reconstruction settings:

\begin{itemize}
    \item \textbf{Full reconstruction:} reconstruct all snapshots $\hat{\mathbf{x}}_{1:T}$
    \item \textbf{Target reconstruction:} reconstruct only the final snapshot $\hat{\mathbf{x}}_T$
\end{itemize}

Target reconstruction reflects practical communication scenarios where only the latest CSI estimate is required.


We evaluate compression performance for latent dimensions $Z \in \{16, 32, 64\}$. Each CSI snapshot is represented by a real-valued vector of dimension $N = 2R = 64$. Thus, the latent dimensions $Z \in \{16,32,64\}$ correspond to progressively weaker compression regimes for single-snapshot models, namely compression ratios of $1/4$, $1/2$, and $1$, respectively. For sequential models, communication cost additionally depends on the number of latent vectors required by the decoder, and is therefore reported separately in Table~\ref{tab:txproxy}.


To assess robustness to incomplete temporal context, we introduce context dropout during inference. In this setting, each past snapshot in the input window is independently removed with probability $p$, while the most recent snapshot is always preserved. This models situations in which part of the past CSI history is unavailable at the encoder input, for example due to missing or delayed measurements. We evaluate $p \in \{0.1, 0.2, 0.3, 0.4\}$.


\begin{table}[t]
\centering
\caption{Transmission cost proxy in number of latent scalars.}
\label{tab:txproxy}
\begin{tabular}{lc}
\toprule
Model & Transmitted latent scalars \\
\midrule
AE / VAE / CVAE & $Z$ \\
TSVAE (full or target) & $8Z$ \\
$k$-MMVAE (full) & $8Z$ \\
$k$-MMVAE (target, $k=3$) & $4Z$ \\
\bottomrule
\end{tabular}
\end{table}

\subsection{Transmission Cost Proxy}

Since latent variables are not quantized or entropy-coded, we approximate the communication cost by the \emph{number of transmitted latent scalars}.

Let $Z$ denote the latent dimension. The transmission cost depends on both $Z$ and the number of latent vectors required by the decoder. Specifically, AE, VAE, and CVAE transmit a single latent vector ($Z$ scalars), whereas TSVAE and $k$-MMVAE (full reconstruction) require one latent per snapshot, resulting in $TZ$ scalars (with $T=8$). For $k$-MMVAE in target reconstruction mode with $k=3$, only the last $k+1=4$ latent vectors are needed, yielding $4Z$ scalars.

The resulting transmission cost proxy is summarized in Table~\ref{tab:txproxy}. Notably, in target reconstruction mode, the cost of $k$-MMVAE depends only on the memory size $k$ and not on the trajectory length $T$, unlike TSVAE where the cost scales linearly with $T$. This property makes $k$-MMVAE particularly suitable for long CSI trajectories.

\subsection{Main Benchmark}

Table~\ref{tab:main_benchmark} presents the test NMSE in dB across three latent dimensions, averaged over three random seeds. The standard deviation assesses stability and robustness across runs.

\begin{table*}[t]
\centering
\caption{Main benchmark results (test NMSE in dB, mean $\pm$ standard deviation over 3 seeds). Best results are shown in bold and second-best results are underlined.}
\label{tab:main_benchmark}
\begin{tabular}{lcccc}
\toprule
Model & Latent scalars transmitted & $Z=16$ & $Z=32$ & $Z=64$ \\
\midrule
AE & $Z$   & $-10.31 \pm 0.05$ & $-16.24 \pm 0.11$ & $\mathbf{-31.98 \pm 0.67}$ \\
VAE & $Z$  & $-10.31 \pm 0.05$ & $-16.11 \pm 0.09$ & $-29.29 \pm 0.33$ \\
CVAE (full) & $Z$   & $-9.62 \pm 0.02$  & $-13.96 \pm 0.05$ & $-23.20 \pm 0.28$ \\
CVAE (target) & $Z$ & $-10.33 \pm 0.03$ & $-15.99 \pm 0.05$ & $-27.76 \pm 0.18$ \\
TSVAE (full) & $8Z$   & $-10.20 \pm 0.01$ & $-15.73 \pm 0.01$ & $-27.85 \pm 0.21$ \\
TSVAE (target) & $8Z$ & $-10.26 \pm 0.03$ & $-15.87 \pm 0.03$ & $-28.64 \pm 0.23$ \\
$k$-MMVAE (k=3, full) & $8Z$ & $\underline{-15.81 \pm 0.03}$ & $\underline{-22.04 \pm 0.08}$ & $-30.18 \pm 0.20$ \\
$k$-MMVAE (k=3, target) & $4Z$ & $\mathbf{-26.90 \pm 0.08}$ & $\mathbf{-28.98 \pm 0.73}$ & $\underline{-31.36 \pm 0.92}$ \\
\bottomrule
\end{tabular}
\end{table*}

\paragraph{Temporal modeling matters most at low and medium latent dimensions}
At $Z=16$ and $Z=32$, the target-version of $k$-MMVAE clearly outperforms all baselines. At $Z=16$, $k$-MMVAE target reaches $-26.90$ dB, whereas the best non-Markov baseline remains near $-10.33$ dB. At $Z=32$, $k$-MMVAE target reaches $-28.98$ dB, while the strongest non-Markov target baseline stays around $-15.99$ dB. This large gap suggests that, at low and medium latent dimensions, spatial reconstruction alone may be insufficient, and that exploiting temporal dependencies can provide significant gains.

\paragraph{Full reconstruction is harder than target reconstruction}
For all sequential models, reconstructing the full trajectory is substantially more difficult than reconstructing only the final snapshot. Nevertheless, $k$-MMVAE full reconstruction consistently outperforms the other full-reconstruction baselines, suggesting that incorporating a Markov prior in the latent space can be beneficial even in the more demanding full-window setting.

\paragraph{At high latent dimension, the advantage of temporal modeling becomes smaller}
At $Z=64$, the best result is achieved by the memoryless AE, with $-31.98$ dB. $k$-MMVAE target remains very competitive at $-31.36$ dB, but the margin relative to the best baseline becomes small. This suggests that, in the high-rate regime, performance may become increasingly dominated by representation capacity rather than temporal structure. 


\subsection{Robustness to Missing Context}

\begin{figure}[t]
    \centering
    \includegraphics[width=\columnwidth]{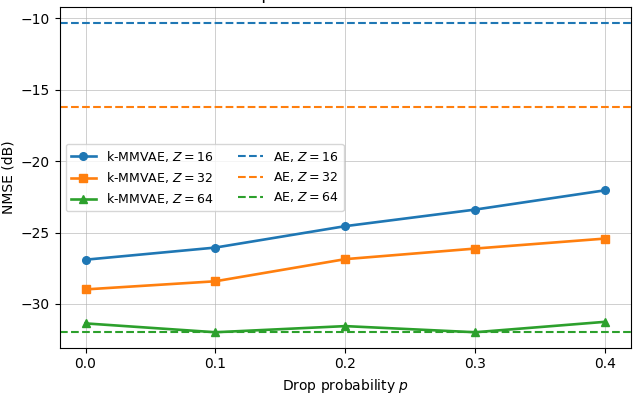}
    \caption{Robustness of $k$-MMVAE target reconstruction to missing temporal context. Solid curves show the NMSE of $k$-MMVAE as $p$ increases. Dashed horizontal lines indicate the corresponding AE performance at the same latent dimension $Z$, serving as memoryless spatial-compression baselines.}
    \label{fig:kmm_drop_target}
\end{figure}

To evaluate whether $k$-MMVAE exploits temporal information rather than merely benefiting from increased model complexity, we study its robustness when past CSI observations are randomly removed at inference time. Fig.~\ref{fig:kmm_drop_target} reports the target-reconstruction results together with the corresponding AE baselines. 

\paragraph{$k$-MMVAE target reconstruction is robust to partial loss of temporal context} At $Z=16$ and $Z=32$, the performance of $k$-MMVAE degrades gradually as the drop probability increases. However, even at $p=0.4$, the model still outperforms the corresponding AE baseline at the same latent dimension. This suggests that $k$-MMVAE continues to benefit from temporal modeling even when part of the past context is unavailable.

\paragraph{The effect of context drop becomes small at high latent dimension} At $Z=64$, $k$-MMVAE target reconstruction remains nearly unchanged across the entire range of drop probabilities and stays close to the AE baseline. This suggests that, in the high-rate regime, performance may be increasingly dominated by representation capacity, making the model less dependent on temporal context.





\section{Conclusion}

This paper investigated neural CSI compression for time-varying massive MIMO channels under limited feedback and proposed a context-aware compression framework based on a $k$-memory Markov variational autoencoder ($k$-MMVAE). The results indicate that incorporating finite temporal context via explicit latent dynamics can significantly improve target CSI reconstruction performance, particularly at low and moderate compression rates, while remaining robust to incomplete temporal observations.

At higher latent dimensions, the advantage of temporal modeling becomes less pronounced, suggesting that performance is increasingly governed by representation capacity. These findings highlight the role of temporal structure as a complementary mechanism to spatial compression in rate-constrained CSI feedback.



\end{document}